\documentclass[twocolumn,showpacs,preprintnumbers,amsmath,amssymb]{revtex4}

\usepackage{graphicx}
\usepackage{dcolumn}
\usepackage{bm}

\newcommand{\frat}[2]{\frac{\textstyle #1}{\textstyle #2}}
\newcommand{\vf}[1]{\mbox{\boldmath $#1$}}
\newcommand{\nomer}[1]{\mbox{$\cal N$\hspace{-.5ex}\raisebox{.3ex}
           {\underline{\tiny 0}$\!$} #1}}

\newcommand{\dmn}[2]{\mbox{$#1\!\cdot\! 10^{#2}\,$}}

\begin{document}

\title{CONDENSATED FERMION SYSTEM IN THE MODEL OF FOUR-QUARK\\
INTERACTION WITH LARGE CORRELATION LENGTH}

\author{S. V. Molodtsov}
 \altaffiliation[Also at ]{
Institute of Theoretical and Experimental Physics, Moscow, RUSSIA}
\affiliation{%
Joint Institute for Nuclear Research, Dubna,
Moscow region, RUSSIA
}%
\author{G. M. Zinovjev}
\affiliation{
Bogolyubov Institute for Theoretical Physics,
National Academy of Sciences of Ukraine, Kiev, UKRAINE
}%

\date{\today}

\begin{abstract}
Studying a model of four-quark interaction with large correlation length
we find out both the features peculiar an unitary fermi gas and the specific
anomalous properties of the fermi systems with a fermion condensate. It is
argued that a possibility of phase transition originated by interface between
the Fermi sphere and fermion condensate appears in such quark systems. The
results obtained could be instrumental for phenomenological applications in
view of our conclusion about approximately the same behavior of the dynamical
characteristics of quark ensembles with different four-quark interaction
forms
in a practical interval of coupling constant.
\end{abstract}

\maketitle

A new form of matter created in ultra-relativistic heavy ion collisions
at RHIC and LHC has been realized as a strongly coupled system in an
anisotropic state with rather unexpected features and certainly making
qualitative
insights to the nature of quark-gluon plasma (QGP). The dynamical evolution
of
this hot and dense system being successfully analysed with the relativistic
viscous hydrodynamics designates another challenge to the QGP (and QCD)
theory
definitely compelling to think of the QGP rather as a liquid than a gas of
quarks
and gluons.
The present letter is devoted to study some aspects of anomalous
thermodynamical state \cite{hs} called a fermion condensate. In particular,
we are interested in analysing relativistic quark ensemble with a specific
form
of four-fermion interaction. These field theory models (QCD like models)
are still most reliable source of qualitative (and quantitative) information
on the transport characteristics of strongly correlated quark systems and the
chiral phase transition between massive hadrons and massless quarks.

The thermodynamical description of the quark ensemble with four-fermion
interaction (generated as it is believed by strong stochastic gluon fields)
is
grounded on the Hamiltonian density
\begin{equation}
\label{1}
{\cal H}=-\bar q~(i{\vf \gamma}{\vf \nabla}+m)~q-j^a_\mu \int d{\vf y}~
\langle A^{a}_\mu A'^{b}_\nu\rangle~j'^b_\nu~,
\end{equation}
where $j^a_\mu=\bar qt^a\gamma_\mu q$ is the quark current, with
operators of the quark fields $q$, $\bar q$, taken in spatial point ${\vf
x}$ (the variables with prime correspond to the ${\vf y}$ point), $m$ is the
current quark mass, $t^a=\lambda^a/2$ is the color gauge group $SU(N_c)$
generators,
$\mu,\nu=0,1,2,3$. The gluon field correlator $\langle A^{a}_\mu
A'^{b}_\nu\rangle$ is taken in
the simplest color singlet form with a time contact interaction (with no
retarding)
\begin{equation}
\label{cor}
\langle A^{a}_\mu A'^{b}_\nu\rangle=G~\delta^{ab}~\delta_{\mu\nu}~F({\vf
x}-{\vf y})~,
\end{equation}
(we do not include the time delta-function in this formula).
This effective Hamiltonian should describe quasi-stationary states
of quark ensemble and it results (in natural way) from the
coarse-grained description of the system (see a derivation of vacuum gluon
fields
in  form of the instanton liquid \cite{MZ}).
Relying on a point-like approximation of correlation function in coordinate
space
we come to the Nambu--Jona-Lasinio model (NJL) \cite{njl}.
The opposite limit of infinite correlation length ($\delta$-function like
behavior in momentum space) is widely used in condensed matter physics
and known as the Keldysh model \cite{kldsh}.
Realizing that for the strongly interacting systems the size
of characteristic vacuum box is found of $\Lambda^{-
1}_{\mbox{\scriptsize{QCD}}}$
order \cite{wearxiv} we could qualitatively expect that both opposite models
lead to practically the same picture of spontaneous chiral symmetry breaking,
color superconductivity and some important features, because a scale of
coupling constant $G$ can be properly tuned by using meson
observables \cite{sscb}, \cite{wemes}.

It is believed (for KKB model it was proved in \cite{wearxiv})
that at strong enough interaction the ground state of
system transforms from trivial vacuum $|0\rangle$ (the vacuum of free
Hamiltonian) to the mixed
state (the quark--anti-quark pairs with opposite momentum with vacuum quantum
numbers),
which is presented as the Bogolyubov trial function (in that way some
separate reference frame
is introduced, and chiral phase becomes fixed)
$$
|\sigma\rangle={\cal{T}}|0\rangle,~
{\cal{T}}=\prod\limits_{ p,s}\exp[\varphi_p~(a^+_{ p,s}b^+_{- p,s}+
a_{ p,s}b_{-p,s})].$$
Here $a^+$, $a$ и $b^+$, $b$ are the quarks creation and annihilation
operators, $a|0\rangle=0$, $b|0\rangle=0$. The dressing transformation
${\cal{T}}$ transmutes the quark operators to the creation and annihilation
operators
of quasiparticles $A={\cal{T}}~a~{\cal{T}}^\dagger$,
$B^+={\cal{T}}~b^+{\cal{T}}^\dagger$.

The thermodynamic properties of the quark ensemble determines by solving
the following problem. It should be found such a statistical operator
\begin{equation}
\label{dm}
\xi=\frat{e^{-\beta ~\hat H_{{\mbox{\scriptsize{app}}}}}}{Z_0}~,
~~Z_0=\mbox{Tr}~\{e^{-\beta ~\hat H_{{\mbox{\scriptsize{app}}}}}\}~~,
\end{equation}
that at fixed mean charge
\begin{equation}
\label{ntot}
\overline{Q}_0=\mbox{Tr} \{\xi ~Q_0\}=
V~\gamma~\int  d \widetilde{\vf p}~(n-\bar n)~,
\end{equation}
($Q_0=\bar q \gamma^0 q$), and fixed mean entropy
\begin{eqnarray}
\label{stot}
\overline{S}&=&-\mbox{Tr} \{\xi~ S\}=\\
&-&V~\gamma~\int d \widetilde {\vf p}~
[n\ln n+(1-n)\ln (1-n)+\nonumber\\
&+&\bar n\ln \bar n+(1-\bar n)\ln (1- n)],\nonumber
\end{eqnarray}
($S=-\ln \xi$), the mean energy of the quark ensemble
$$E=\mbox{Tr} \{\xi~H\}~,$$
($H=\int d{\vf x}~ {\cal H}$) would be minimal. In other words we are
interested in finding
the minimum of the following functional
\begin{equation}
\label{fun}
\Omega=E-\mu~\overline{Q}_0 -T~\overline{S}~,
\end{equation}
where $\mu$ and $T$ denote the Lagrangian multiplier for chemical
potential and the temperature respectively ($\beta=T^{-1}$). V is a
volume in which the system is enclosed, $d \widetilde{\vf p}=d {\vf
p}/(2\pi)^3$,
$\gamma=2 N_c$ (in the case of quarks of a few flavors $\gamma=2 N_c N_f$,
where $N_f$ is the number of
flavors), $n=\mbox{Tr} \{\xi A^+ A\}$, $\bar n=\mbox{Tr} \{\xi B^+ B\}$ are
the
components of corresponding density matrix.
\begin{figure}
\includegraphics[width=0.3\textwidth]{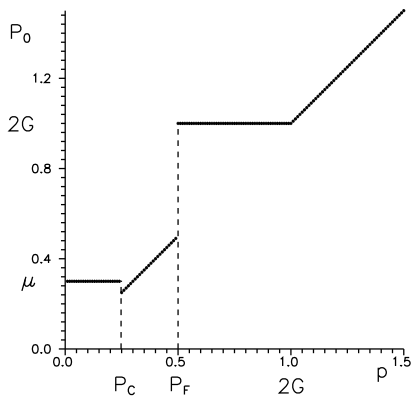}
\caption{ Quark energy as a function of momentum for the first (see the text)
solution.}
\label{f3}
\end{figure}

We restrict ourselves by considering the Bogolyubov--Hartree--Fock
approximation in which the statistical operator is constructed on the basis
of approximated
effective Hamiltonian $H_{{\mbox{\scriptsize{app}}}}$ quadratic in creation
and annihilation
operators of quasi-particles acting in the corresponding Fock space with a
vacuum
state $|\sigma\rangle$. The average specific energy per quark $w=E/(V\gamma)$
is given
\cite{MZ2} by the following form
\begin{eqnarray}
\label{w}
w&=&\int d \widetilde{\vf p}~p_0-\int d \widetilde{\vf p}~
(1-n-\bar n)~p_0~\cos\theta-
\nonumber\\[-.2cm]
\\ [-.25cm]
&-&\frac12~\int d \widetilde{\vf p}~(1-n-\bar n)
\sin \left(\theta-\theta_m\right)~M({\vf p})~,\nonumber
\end{eqnarray}
where
$$
M({\vf p})=2G\int d \widetilde{\vf q}~(1-n'-\bar n')~
\sin \left(\theta'-\theta'_m\right)~F({\vf p}+{\vf q})~,$$
$\theta=2\varphi$, $p_0=({\vf p}^2+m^2)^{1/2}$, the primed variables,
here and below, correspond to the integration over momentum ${\vf q}$.
The auxiliary angle $\theta_m$ is determined from the relation: $\sin
\theta_m=m/p_0$.
The first term in Eq. (\ref{w}) is introduced in view of normalization in
order to have the
zero ground state energy when an interaction is switched off. This constant
would be
inessential in what follows and can  be safely omitted. However, it should be
kept in
mind that it will appear further as a regularizer in singular expressions if
they occur.
\begin{figure}
\includegraphics[width=0.3\textwidth]{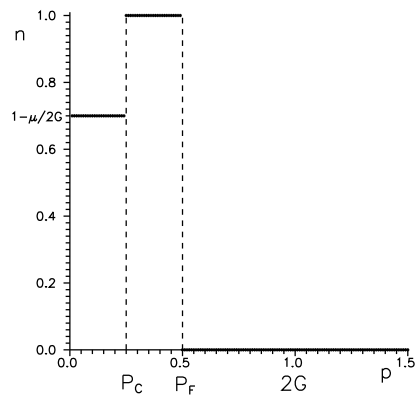}
\caption{
Quark ensemble density as a function of the momentum for the first (see the
text) solution.
 Fermion condensate solution is followed by the Fermi sphere,
 and then by the vacuum solution.}
\label{f4}
\end{figure}

 For the delta-like potential in coordinate space (NJL model)
the expression (\ref{w}) diverges and to obtain the reasonable results the
cut-off upper
limit over momentum integration $\Lambda$ is introduced, which along with the
coupling constant $G$ and current quark mass $m$ is one of the tuning model
parameter. Below we use
one of the standard parameter sets for the NJL model \cite{5}: $\Lambda=631$
MeV,
$G\Lambda^2/(2\pi^2)\approx 1.3$, $m=5$ MeV while the KKB model
parameters are chosen in such a way that for the same current masses the
quark dynamical masses in both NJL and KKB models coincide at vanishing quark
momentum.

\begin{figure}
\includegraphics[width=0.3\textwidth]{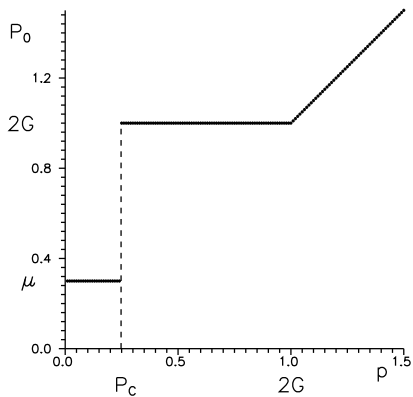}
\caption{Quark energy as a function of momentum for the second (see the text)
solution.}
\label{f5}
\end{figure}
Using the extremal properties the functional Eq. (\ref{w}) can be
transformed to the form (see \cite{MZ2})
\begin{eqnarray}
\label{w2}
w&=&\int d \widetilde{\vf p}~p_0-\int d \widetilde{\vf p}~
(1-n-\bar n)~P_0+
\nonumber\\[-.2cm]
\\ [-.25cm]
&+&\frac{1}{4G}~\int d \widetilde{\vf p}d \widetilde{\vf q}~~
F({\vf p}+{\vf q})~\widetilde M({\vf p})\widetilde M({\vf q})~,\nonumber
\end{eqnarray}
where $P_0=[{\vf p}^2+M_q^2({\vf p})]^{1/2}$ is the energy of quark
quasi-particle with a quark dynamical mass
\begin{equation}
\label{9}
M_q({\vf p})=m+M({\vf p})=m+\int d \widetilde{\vf q}~F({\vf p}+{\vf
q})~\widetilde M({\vf q})~.
\end{equation}
Below we omit the arguments of corresponding functions for the mass and
quasiparticle energy. Varying the functional (\ref{w2}) with respect to
the density of induced quasi-particle mass $\widetilde M$ (in such a form it
is convenient to
calculate variational derivatives{\footnote{If one takes the quark dynamical
mass $M_q$ as a
basic variable, then it is seen from Eq. (\ref{9}) that it is difficult to
formulate an inverse transformation from $M_q$ to $\widetilde M$ suitable for
handling.}})
we obtain the following equation for dynamical quark mass
\begin{equation}
\label{10}
M_q({\vf p})=m+2G\int d \widetilde{\vf q}~(1-n'-\bar
n')~\frac{M'_q}{P'_0}~F({\vf p}+{\vf q}),
\end{equation}
which exactly corresponds to the mean field approximation. In particular,
under normal condition ($T=0$, $\mu=0$) the quark dynamical mass in NJL model
is $M_q\sim 340$
MeV, while the quark dynamical mass of the KKB model is determined by the
equation
\begin{equation}
\label{mkeld}
M({\vf p})=2 G~\frac{M_q({\vf p})}{P_0}~.
\end{equation}
In practice it is convenient to deal with the inverse function $p(M_q)$.
In particular, in the chiral limit $M_q=(4G^2-{\vf p}^2)^{1/2}$ for $|{\vf
p}|<2G$, and $M_q=0$ at
$|{\vf p}|>2G$. Then, the quark states with momenta $|{\vf p}|<2G$ are
degenerate in energy $P_0=2G$.

\begin{figure}
\includegraphics[width=0.3\textwidth]{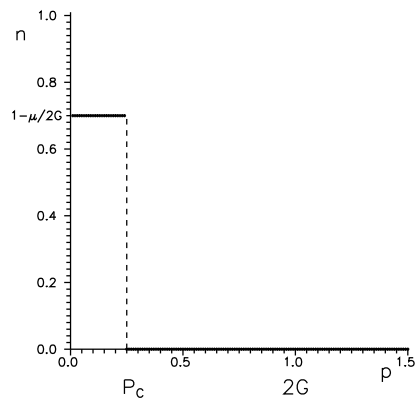}
\caption{
Quark ensemble density as a function of the momentum  for the second (see the
text) solution.
 $P_{\tiny C}$ value separates a fermion condensate and vacuum contributions.
}
\label{f6}
\end{figure}
It was suggested in \cite{hs} that, besides the standard Fermi
distribution, the anomalous states (fermion condensate) are possible. We
study them here with the KKB model
example discussing, first, the situation of zero temperature. We need to find
a minimum of the
functional (\ref{fun}) at a fixed mean charge (baryon number)
$\overline {\cal N}=\gamma\int d \widetilde {\vf p}~ n({\vf p})$,
and mean entropy $\overline {\cal S}=\gamma\int d \widetilde {\vf p}~
s({\vf p})$. By varying the quark dynamical mass $M_q$ and density $n$ we
obtain the system of equations
\begin{eqnarray}
\label{pp1}
&&-(1-n)~\frac{M_q}{P_0}+\frac{M}{2G}=0~,\nonumber\\[-.2cm]\\[-.25cm]
&&
P_0-\mu-T \ln(n^{-1}-1)=0~.\nonumber
\end{eqnarray}
For the fermi-condensate it is proposed to make use the second equation
of the system (\ref{pp1}) and search for a solution in the form
\begin{eqnarray}
\label{pp2}
&&T\equiv 0~,\nonumber\\[-.2cm]\\ [-.25cm]
&&
P_0=\mu~,\nonumber
\end{eqnarray}
degenerate in energy (as was shown above a similar behavior is
demonstrated by the KKB model).
Then for the quark dynamical mass we have
\begin{equation}
\label{pp3}
M_q=\pm(\mu^2-{\vf p}^2)^{1/2}~,~
\end{equation}
and $|{\vf p}|<\mu$ if one is interested in the real solutions only.
Besides, there exist, of course, a standard solution which is considered to
be an asymptotic form
of the Fermi distribution at $T\to 0$
$$n=\frac{1}{e^{\beta(P_0-\mu)}+1}~,$$
with $P_0=[{\vf p}^2+M_q^2({\vf p})]^{1/2}$. The quark dynamical mass is
defined by the following relations
\begin{eqnarray}
&&\frac{M_q}{P_0}=\frac{M}{2G}~,~~n=0~,~~|{\vf p}|>P_{\tiny
F}~,\nonumber\\
[-.2cm]\\ [-.25cm]
&&M=0~,~~~~n=1~,~~|{\vf p}|<P_{\mbox{\tiny{F}}}~.\nonumber
\end{eqnarray}
By definition, the condensate density satisfies inequalities $0<n<1$.
From the first equation of the system (\ref{pp1}) we find two possible
density distributions
\begin{equation}
\label{p4}
n_\pm=1-\frac{\mu}{2G}\pm\frac{\mu}{2G}\frac{m}
{(\mu^2-{\vf p}^2)^{1/2}}~.
\end{equation}
It is interesting to note the peak in the density $n_+$ at
$|{\vf p}|\sim\mu$, but at the same time the quark density cannot exceed $1$.
 For the second solution $n_-$
another constraint $n>0$ is valid. Going to find the domain of the solution
$n_+$ applicability we
define the momentum $p_+$ in a  way that $n_+=1$, i.e. $p_+=(\mu^2-
m^2)^{1/2}$. It is obvious that
the momentum $p_+$ is separated from the value $\mu$, where the density $n_+$
is singular, by a
constant value that is defined by the quark current mass. Defining the region
of the second
solution $n_-$ applicability we define the momentum $p_-$ to have $n_-=0$,
i.e. $p_-=[\mu^2-
m^2\mu^2/(2G-\mu)^2]^{1/2}$. In contrast to the momentum $p_+$ the limiting
momentum found is movable
with respect to the value $\mu$, at $\mu\to 0$ we have $p_-\to \mu$. When
$\mu=G$, the momenta
$p_+$ and $p_-$ coincide ($p_+=p_-$). If $\mu=\mu_-=2G-m$ then $p_-$ goes to
zero. For larger
$\mu>\mu_-$ the second branch of solution, $n_-$, disappears. Summarizing, we
may
conclude that it is possible to have the situations in which there exist two
solutions for
the condensate within the interval of momenta. Then beyond this region an
interval can be
situated where only one solution exists either, $n_+$ or $n_-$ depending on
the relation
between momenta $p_+$ and $p_-$. And, finally, beyond this latter interval
only the
solution with a standard Fermi distribution can exist.
\begin{figure}
\includegraphics[width=0.3\textwidth]{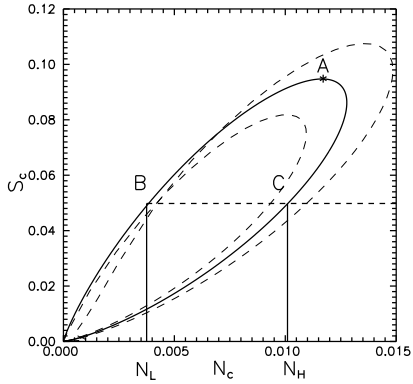}
\caption{The entropy density (per fm$^3$) as a function of quark ensemble
density (per fm$^3$) for a fermi condensate state. The solid curve is
obtained in the chiral limit, which serves as an example for examining an
issue of quark ensemble density increase at fixed mean entropy.}
\label{f14}
\end{figure}

Further analysis can be carried out in the chiral limit only. In this
case both branches $n_+$ and $n_-$ get merged and the condensate density does
already not depend
on the quark momentum
$$n=n_+=n_-=1-\frac{\mu}{2G}~.$$
It is seen that the condensate solution is possible only at $\mu<2G$.
Then $M_q=M=(\mu^2-{\vf p}^2)^{1/2}$, and here the real solution exists only
within
the interval $0\leq |{\vf p}|\leq \mu$. In addition to the condensate
solution the
standard one is also possible\\
$\begin{array}{ll}
M_q=M=[(2G)^2-{\vf p}^2]^{1/2},~& n=0~,~~|{\vf p}|>P_{\mbox{\tiny{F}}} \\
M=0~,&n=1~,~~|{\vf p}|\leq P_{\mbox{\tiny{F}}}~.
\end{array}
$

\begin{figure}
\includegraphics[width=0.3\textwidth]{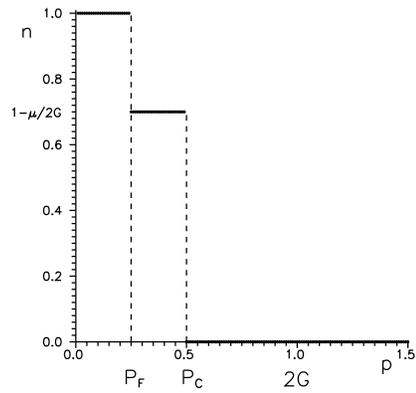}
\caption{
Quark ensemble density as a function of the momentum for the third (see the
text) solution.
The fermion condensate and Fermi sphere are rearranged here comparing to the
first solution,
see Fig. \ref{f4}.
}
\label{f8}
\end{figure}

It is easy to understand that the general solution can be obtained by
combining
the standard solutions of the Fermi step
and fermion condensate at different intervals of momentum axis.
We consider a few such possibilities.
Figs. \ref{f3}, \ref{f4} demonstrate the quark energy and quark ensemble
density
as the functions of momentum. We place
the solution with a Fermi condensate into the interval $[0,P_{\tiny C}]$
localizing the Fermi sphere in the interval $[P_{\tiny C}, P_{\tiny F}]$ and
vacuum solution is placed behind the Fermi momentum $P_{\tiny F}$.
According to definition we take here $P_{\tiny C}<P_{\tiny F}$,
$\mu\ge P_{\tiny C}$ and call such functions as the first solution.
Detaching the solution without Fermi sphere we call it as the second
solution.
Figs. \ref{f5} and \ref{f6} show the corresponding quark energy and quark
ensemble
density. Thus, there is a fermion condensate that is followed with the vacuum
solution along the momentum axis and then $\mu\ge P_{\tiny C}$.
It could be convenient to characterize the solutions with the
dimensionless variables $x=\mu/(2G)$, $y=P_{\tiny F}/(2G)$, $z=P_{\tiny
C}/(2G)$.
The mean entropy density and the particle number density in the fermion
condensate
 for the second solution are given by:
\begin{eqnarray}
\label{pp5}
&&{\cal S}_2=-\frac{\gamma}{6\pi^2}~z^3\left[(1-x)\ln (1-x)+x \ln
x\right]~(2G)^3~,
\nonumber\\[-.1cm]\\ [-.1cm]
&&{\cal N}_2=\frac{\gamma}{6\pi^2}~z^3 (1-x)~(2G)^3~.\nonumber
\end{eqnarray}
When $\mu= P_{\tiny C}$ the fermion condensate contains
maximally possible number of states.
Fig. \ref{f14} explores the entropy density (over fm$^3$)
of the Fermion condensate as function of baryon density
${\cal N}=Q_0/(3V)$, and we specify it as $2G=300$ MeV.
The solid oval line is obtained in the chiral limit. We present it in
physical units
in order to estimate the order of magnitude of the characteristics.
But in what follows we characterize the entropy, quark ensemble density and
ensemble
energy in dimensionless variables (with the corresponding powers of
coefficient $2G$).
Fig. \ref{f14} shows also the entropy density for the quark ensemble with
current quark
mass $m=5$ MeV. $S_+$ (large dashed oval), and $S_-$ (small dashed
oval) was obtained by making use the distributions $n_+$ and $n_-$
correspondingly. The maximal condensate density is achieved at $x_N=3/4$,
$N_c\approx \dmn{3.56}{-3}(2G)^3$,
maximal entropy occurs at $x_S\approx 0.84$,  $S_c\approx \dmn{2.64}{-
2}(2G)^3$ $(N_f=1)$.
The fermion condensate states with $P_{\tiny C}<\mu$ populate an oval
interior.
The energy density of the second solution is calculated from Eq. (\ref{w2})
(where the integration is extended up to the boundary momentum $2G$ only
because the large values of chemical potential and momentum $P_{\tiny C}$,
$P_{\tiny F}$
are unrealistic)  in the following form
\begin{equation}
\label{e2}
{\cal E}_2=-\frat{\gamma}{4\pi^2}\left(\frat{8}{15}-
\frat{z^3}{3}+\frat{z^3x^2}{3}\right)~(2G)^4~.
\end{equation}
\begin{figure}
\includegraphics[width=0.3\textwidth]{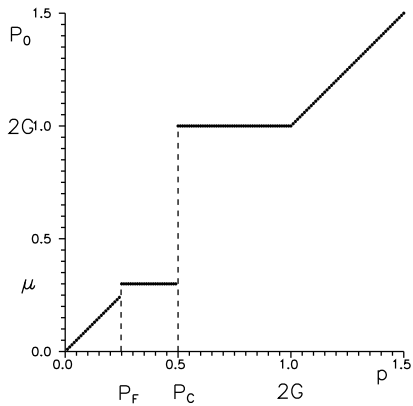}
\caption{
Quark energy as a function of momentum for the third (see the text)
solution.}
\label{f7}
\end{figure}
The solutions obtained could be interpreted as nontrivial continuation of
a standard procedure of filling in the Fermi sphere (in that case
chemical potential is equal to or exceeds the
dynamical quark mass by definition) up to the situation when the chemical
potential values
become smaller than $M_q$.
Figs. \ref{f7} and \ref{f8} show the third solution with
  $P_{\tiny C}> P_{\tiny F}$, $\mu\ge P_{\tiny C}$.
It looks like the first solution, but with momenta separating the Fermi
sphere from the
fermion condensate rearranged. Here we do not discuss  the solution with
Fermi sphere only
(without fermion condensate) because the entropy of this state is equal to
zero.

Then average particle number density, average entropy density and  average
energy density
 for the first solution are the following:
\begin{eqnarray}
\label{e1}
&&{\cal N}_1=\frac{\gamma}{6\pi^2}~(y^3-z^3 x)~(2G)^3~,\\
&&{\cal S}_1=-\frac{\gamma}{6\pi^2}~z^3\left[(1-x)\ln (1-x)+x \ln
x\right]~(2G)^3~,\nonumber\\
&&{\cal E}_1=-\frat{\gamma}{4\pi^2}\left(\frat{8}{15}-\frat{y^3}{3}+
\frat{z^3x^2}{3}+\frat{z^5}{5}-\frat{y^5}{5}\right)~(2G)^4~~.\nonumber
\end{eqnarray}
Similar quantities for the  third solution look like:
\begin{eqnarray}
\label{e3}
&&{\cal N}_3=\frac{\gamma}{6\pi^2}~(z^3(1-x)+y^3 x)~(2G)^3~,\\
&&{\cal S}_3=-\frac{\gamma}{6\pi^2}~(z^3-y^3)\left[(1-x)\ln (1-x)+x \ln
x\right]~(2G)^3~,\nonumber\\
&&{\cal E}_3=-\frat{\gamma}{4\pi^2}\left(\frat{8}{15}-\frat{z^3}{3}+
\frat{(z^3-y^3)x^2}{3}-\frat{y^5}{5}\right)~(2G)^4~~.\nonumber
\end{eqnarray}

In order to find the minimal energy at fixed average entropy, and the average
quark ensemble density
we analize auxiliary function $[-(1-x)\ln (1-x)-x \ln x]$.
This function develops the maximal value $\ln 2$ at the point $x=0.5$ and
at the point $x=0$ and $x=1$ it possesses the minimum value equal to zero.
There are two roots of equation $[-(1-x)\ln (1-x)-x \ln x]=c$ for $0<c<\ln 2$
and due to symmetry arguments, the second root for $x>0.5$, $x_2=1-x_1$ is
obviously determined by
the root $x_1$ for $x<0.5$. As the "reference"\ solution we consider the
second one, because it is
simpler to realize a searching algorithm considering the relations (\ref{e2})
as a system of
equations for $x$ and $z$. (Then the similar analysis could be fruitfully to
the first and third
solutions.) At a fixed entropy $0<{\cal S}<{\cal S}_{{\mbox{max}}}$
the condensate solution is located in the interval $B$, $C$,
see Fig. \ref{f14}.
Making use a standard method of interval bisection we are searching
particular value of auxiliary
function---$c$. This value $c$ is fixed by a constraint to have
$z^3$ (defined by the corresponding $x$ and running value ${\cal N}_{\tiny
C}$)
from the first line of Eq. (\ref{e2})
in coincident (within required precision) with $z^3$ from the second line.
(It is clear that both branches of the auxiliary function mentioned above
should be taken into account.)
Obviously, the similar construction (Eq. (\ref{e2}))
could be applied for the first solution analysis,
but should be added by the contributions of the states falling into the Fermi
sphere
$${\cal N}_1=\frac{\gamma}{6\pi^2}~[y^3-z^3 +z^3(1-x)]~(2G)^3~.$$

Actually, it is more convenient to realize that in two steps.
First, we define $x$ and $z$ at ${\cal N}_{\tiny L}\leq{\cal N}_{\tiny C}
\leq{\cal N}_{\tiny R}$,
then at the second step $ 0\leq {\cal N}_{\tiny F} \leq{\cal N}_{\tiny B}-
{\cal N}_{\tiny C}$
where ${\cal N}_{\tiny B}$ denotes a maximal quark ensemble density and
 $${\cal N}_{\tiny F}=\frac{\gamma}{6\pi^2}~(y^3-z^3)~,$$
we determine $y$.
The total density of the quark ensemble is:
 $${\cal N}={\cal N}_{\tiny C}+{\cal N}_{\tiny F}~.$$
 Similarly, one can deal with the third solution.
It is seen from Eq. (\ref{e3})
 that now in the "reference"\ algorithm  instead of $z^3$ the of $z^3-y^3$
appears
 $${\cal N}_3 =\frac{\gamma}{6\pi^2}~[y^3 (z^3-y^3)(1-x)]~(2G)^3~,$$
where the state density of the Fermi condensate is
 $$ {\cal N}_{\tiny F}=\frac{\gamma}{6\pi^2}~y^3~.$$
Then the total ensemble density is defined as ${\cal N}={\cal N}_{\tiny
C}+{\cal N}_{\tiny F}$.
We passed all the steps for the first solution similarly to the analysis done
above.

\begin{figure}
\includegraphics[width=0.35\textwidth]{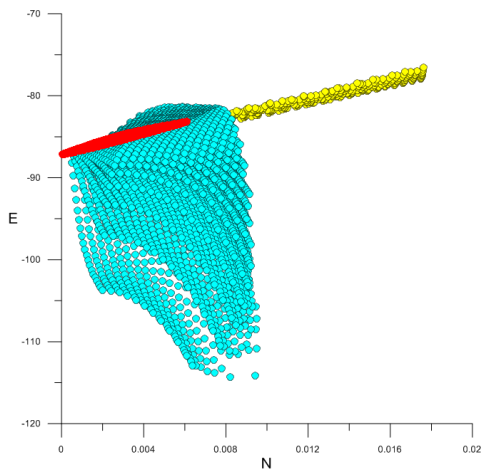}
\caption{ Energy density as a function of the quark density of the ensemble
for the three solutions.}
\label{f9}
\end{figure}

Further, proceeding to the qualitative analysis
we are based on the knowledge of the state energies
as a function of quark/baryon ensemble density
(in fm$^3$) (baryon density is in factor three smaller than
the quark one) putting those on the ${\cal E}$---${\cal N}$ plane, Fig.
\ref{f9}.
(The thorough analysis supposes a consideration of envelope of the curves.)
 The red region in this Fig. \ref{f9} corresponds to the second solution,
and shows the quark ensemble state at low densities, where (at non-zero
entropy) the contribution
of the Fermi sphere is significantly suppressed.
At the densities ${\cal N}\sim\dmn{4}{-4}$---$\dmn{5}{-4}$ (in dimensionless
units)
the contribution of the states filling in the Fermi sphere,
which are described by third solution (blue dots in Fig. \ref{f9}), starts to
increase.

Characteristic values of the other parameters are the following:
$x\sim 0.25$, $z\sim 0.18$ (for the second solution) and
$x\sim 1$, $y\sim 0.15$, $z\sim 0.4$ (for the third solution). The density of
fermion condensate is
estimated to be high $n\sim 0.7$ in the second solution and for the third
solution it is lower,
however the process of filling in the Fermi sphere provides quite noticeable
impact.
It looks like that at such densities the quarks spill over from the fermion
condensate into the Fermi sphere. At further increase of ensemble density the
process
of the Fermi sphere filling in with the fermion condensate
is described by the first solution (yellow dots in Fig. \ref{f9}).
Characteristic ensemble densities when the transition
from the state in which the fermion condensate is
available at large momenta (see Fig. \ref{f8})
 to the state where the fermion condensate exist at small momenta (see Fig.
\ref{f4}) are estimated
as ${\cal N}\sim \dmn{8}{-3}$ with
 $x\sim 1$, $y\sim 0.15$, $z\sim 0.4$ (for the third solution) and
$x\sim 0.2$, $y\sim 0.4$, $z\sim 0.2$ (for the first solution).

It means that in the region of large momenta the low density fragment
of fermion condensate (resulting from the third solution) spills over
 into low momenta region (resulting from the first solution)
reaches remarkable density $n\sim 0.8$.
This rearrangement of quark ensemble behavior is accompanied by relatively
high
energy release (absorption) of order about 20 MeV/fm$^3$ (with $2G=$300 MeV).
Analysis of a general solution including an alternation of different
fragments of the fermion condensate and the Fermi sphere is quite
complicated,
and it is a reason why we are focused only on the analysis done above.
It was also mentioned that spontaneous breaking of chiral symmetry and other
possible phase transitions take place in a similar way in both NJL
and KKB models. It is easy to show that a similar situation happens to
a fermion condensation considered here, albeit we
argued it dealing with the KKB model only. The visible difference
is that instead of a constant quark energy characteristic for the KKB
model (easily seen in Figs.) the parabolic structures appear which correspond
to the constant quark mass (just an approximation in which the NJL model is
valid).
However we are not calculating it here and refer to the result of Ref.
 \cite{sscb}.

Now turning to the situation of final temperature we restrict ourselves
to analyzing the solutions in the chiral limit only
and keeping in mind that the limit $T\to 0$, as we see,
leads to an essentially singular point.
Then an explicit
dependence on the momentum is absent that makes of
course a considerable convenience for analysis. Here it is necessary to take
into account the
anti-quark contribution resulting in the system (\ref{pp1}) to get the form
\begin{eqnarray}
\label{pp7}
&&-(1-n-\bar n)~\frac{M_q}{P_0}+\frac{M}{2G}=0~,\nonumber\\[-.2cm]
\\ [-.25cm]
&&
P_0-\mu-T \ln(n^{-1}-1)=0~,\nonumber\\
&&
P_0+\mu-T \ln(\bar n^{-1}-1)=0~.\nonumber
\end{eqnarray}
In the chiral limit we have
\begin{equation}
\label{pp8}
1-n-\bar n=\frac{P_0}{2G}~.
\end{equation}

\begin{figure}
\includegraphics[width=0.3\textwidth]{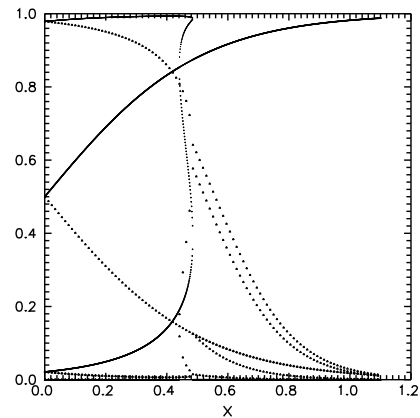}
\caption{Five branches of solutions to Eqs. (\ref{pp11}), (\ref{pp12})
for the quark and anti-quark densities (which correspond to the curves
with lower density of dots) with parameter $y=0.25$ as a function of
the parameter $x$.}
\label{f15}
\end{figure}
\begin{figure}
\includegraphics[width=0.3\textwidth]{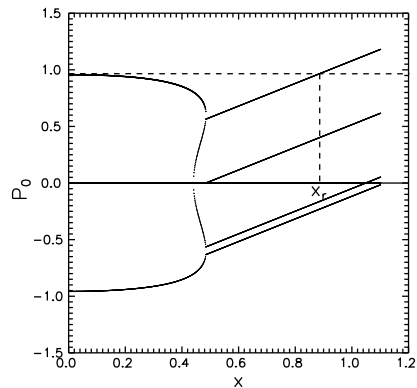}
\caption{The quark energy $P_0$ for solutions presented in Fig.
\ref{f15}. One may see the roots with negative energy.
}
\label{f16}
\end{figure}

It is curious to note already at this point that now it becomes possible
to have the situations with negative quark energy,
i.e. formally it corresponds to the bound state of a quasi-
particle. The energy of quasi-particle with non-zero
dynamical mass is constrained by the inequalities $-2G<P_0<2G$.
For quarks with higher energies, $P_0>2G$, the first equation
of the system leads to the trivial solution with zero
quark dynamical mass $M_q=M=0$. From the second equation (\ref{pp7}) we have
\begin{equation}
\label{pp9}
P_0=\mu+T \ln(n^{-1}-1)~.
\end{equation}
For convenience, we introduce another dimensionless variables
$x=\mu/(2G)$, $y=T/(2G)$ and substituting the energy
in Eq. (\ref{pp8}) we obtain
\begin{equation}
\label{pp11}
1-n-\bar n=x+y \ln(n^{-1}-1)~.
\end{equation}
Linking up the third equation of the system (\ref{pp7}) we can explicitly
find the density of anti-quarks as
\begin{equation}
\label{pp12}
\bar n=\left(e^{2x/y+\ln (n^{-1}-1)}+1\right)^{-1}~,
\end{equation}
and putting it in Eq. (\ref{pp11}) allows us to derive a final
transcendental equation to be used in computations.
To give an illustration we make use the dimensionless variables, i.e. all the
characteristics to be
divided by the corresponding powers of parameter $2G$. Fig. \ref{f15}
displays five solutions to
Eqs. (\ref{pp11}), (\ref{pp12}) for the quark and anti-quark densities
(the dots on the curves are sparser) with parameter $y=0.25$ as a function of
parameter $x$. The
number of branches of the transcendental equations system  evolves with a
change of parameter $y$.
The chosen value $y=0.25$ corresponds to the most abundant number of roots
(remember that at zero
temperature and beyond the chiral limit there were only two branches of
solutions for the density).
It is also interesting to mention that there appear the states with higher
anti-quark density at
rather moderate temperatures. Fig. \ref{f16} illustrates the mentioned
possibility of having the
solutions with negative quark energy which are exactly due to the
considerable anti-quark
contribution.
(An observed value of the charge density is given by the difference of two
large numbers $n$ and
$\bar n$.) A nontrivial solutions for the condensate should satisfy the
energy constraint
$|P_0|<2G$. The figure also shows the straight line $P_0=2G$. Its
intersection point with the curve
gives a limiting value of the chemical potential, at which the quark
condensation (generation of the
quark dynamical mass) for the considered branch of solution is still
possible. In the figure this
point is denoted as $x_r$.
Now we define some integral characteristics of the quark ensemble. For
example, the mean charge and entropy densities look like
\begin{eqnarray}
{\cal Q}_0&=&\gamma\int d \widetilde {\vf p}~(n-\bar n)~,\nonumber\\
{\cal S}&=&\gamma\int d \widetilde {\vf p}~(s+\bar s)~.\nonumber
\end{eqnarray}
By definition, the energy is expressed by the quark dynamical mass as \\
$P_0=\left\{
\begin{array}{l}
\pm\left({\vf p}^2+M^2\right)^{1/2}~,~~ |P_0|\leq 2G~,
\\
\pm |{\vf p}|~,~~ |P_0|>2G~.
\end{array}
\right.
$
\\
Here the quark momentum $|{\vf p}|$ is running within the interval from
$0$ up to $|P_0|$. Then we have
\begin{eqnarray}
{\cal Q}_0&=&\frac{\gamma}{6\pi^2}~|P_0|^3~(n-\bar n)~,\nonumber\\
{\cal S}&=&\frac{\gamma}{6\pi^2}~|P_0|^3~(s+\bar s)~,\nonumber
\end{eqnarray}
with $P_0=2G~[x+y\ln(n^{-1}-1)]$.

\begin{figure}
\includegraphics[width=0.3\textwidth]{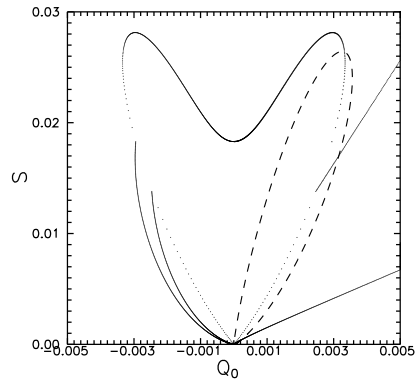}
\caption{The entropy density as function of the density of baryon charge,
at temperature $y=0.25$. Dashed line corresponds to the situation of zero
temperature, see Fig. \ref{f14}.}
\label{f19}
\end{figure}

Fig. \ref{f19} shows the entropy as a function of charge density at
temperature $y=0.25$ when the largest number of solutions to Eq. (\ref{pp7})
is revealed in the
chiral limit. In order to compare the dashed line demonstrates an oval
obtained at zero temperature
which was discussed above. Its evolution with temperature increasing can
clearly be traced.
The changes take place mostly due to the contribution of anti-quarks and
are seen to affect the left hand branch of an oval. The right hand part of
oval stays sort of more
conservative. In this sense it is possible to say that with an increasing
temperature of ensemble
there exist some temperature window where substantial asymmetry in
quark/anti-quark ensemble is
manifested. With these amazing results, we limit our analysis in the present
paper. In order to
examine the state of ensemble as a function of mean entropy and mean charge
in a way similar to what
was done in the situation of zero temperature, it is necessary to analyse
more carefully the chiral
limit, $m\to 0$, of solutions to the equation system (\ref{pp7}). We
demonstrate the states of
ensemble with the fermi-condensate at the temperature approaching the
absolute zero may occupy the
whole semi-plane bounded at the $S-N$ plane by maximal value of accessible
entropy $S<S_c$. (In fact,
this result could be considered as another example of the Nernst 'heat
theorem' breakdown that has
been predicted for strongly correlated fermi-systems of condensed matter
physics \cite{hs},
\cite{amsti}.

Here, it is worth to remind one remarkable fact for those who is interested
in further development of
such an approach. The models of similar Hamiltonian forms were (and are)
widely used in the physics
of condensed matter and nuclear physics while dealing with the ensembles of
finite particle numbers.
They are exactly integrable \cite{richard}, \cite{gaudin} and well understood
in the framework of
conformal theory \cite{sierra}. It encourages us to construct a field theory
model with an increasing
correlation length to trace back, in a sense, field theory origin of the BCS-
type phenomena and,
perhaps, to develop a fresh look at the deconfinement conception.

Summarizing we would like to emphasize that our unexpected point in this
paper concerns the statement
about the possible rearrangement of the quark ensemble
with energy release (absorbtion) about 20 MeV/fm$^3$ (for $2G=$300 MeV) at
its density increasing.
It seems this rearrangement of quark ensemble could be instrumental in the
astrophysical applications,
in particular, to study the problem of Supernova outburst \cite{nad}.

These solutions to the system of thermodynamic equations are quite
different from the standard ones because of very high ensemble density that
in considerable extent is
caused by significant contribution of anti-quarks. Our ensemble displays the
features which are shared
by, for example, the unitary Fermi gas and could be pretty universal. The
latter is considered as one
of the strongest correlated systems in the nature because it saturates the
unitary bound for the
$s-wave$ cross section and develops, as known, the features similar to QGP.
We hope to return to
discussing these problems in more general context of quantum phase
transitions and anomalous behavior
of Fermi-systems \cite{amsti} in future, and now concluding we would like to
mention that going to
perform a similar analysis of Fermi condensate in the NJL-model we have to
deal with the non-local
formulations. The remarkable advantage of our analysis here (which can be
quite practical in studying an
origin of turbulence in QGP) is the locality of interaction in the momentum
space.

ACKNOWLEDGMENTS

Authors are deeply indebted to K. A. Bugaev, I. M. Dremin, V. V. Goloviznin,
E.-M. Ilgenfritz, A. V. Leonidov, D. K. Nadezhin, S. N. Nedelko,
 V. V. Skalozub, A. M. Snigirev and many other colleagues for numerous
fruitful discussions.
The work was partially supported by the State Fund for Fundamental Research
of Ukraine,
Grant \nomer{Ph58/04}.


\end{document}